\documentclass[aps,twocolumn,a4paper,showkeys,showpacs, superscriptaddress]{revtex4-1}
\usepackage{graphicx}
\usepackage{color}

\begin{document}

\title{$A$-site driven ferroelectricity in strained ferromagnetic La$_2$NiMnO$_6$ thin films}
\author{R. Takahashi}
\email[]{rtaka@issp.u-tokyo.ac.jp}
\affiliation{Institute for Solid State Physics, University of Tokyo,
             5-1-5, Kashiwanoha, Kashiwa, Chiba 277-8581, Japan}

\author{I. Ohkubo}
\affiliation{Department of Applied Chemistry, School of Engineering,
 University of Tokyo, 7-3-1 Hongo, Tokyo 113-8656, Japan}
\affiliation{National Institute for Materials Science, Tsukuba 305-0044, Japan}
              
\author{K. Yamauchi}
\affiliation{Institute of Scientific and Industrial Research, Osaka University, Ibaraki, Osaka 567-0047, Japan}

\author{M. Kitamura}
\author{Y. Sakurai}
\affiliation{Department of Applied Chemistry, School of Engineering,
 University of Tokyo, 7-3-1 Hongo, Tokyo 113-8656, Japan}

\author{M. Oshima}
\affiliation{Department of Applied Chemistry, School of Engineering,
 University of Tokyo, 7-3-1 Hongo, Tokyo 113-8656, Japan}
 \affiliation{Synchrotron Radiation Research Organization,
  University of Tokyo, Tokyo 113-8656, Japan}

\author{T. Oguchi}
\affiliation{Institute of Scientific and Industrial Research, Osaka University, Ibaraki, Osaka 567-0047, Japan}
\affiliation{JST-CREST, Kawaguchi, Saitama, 332-0012, Japan}

\author{Y. Cho}
\affiliation{Research Institute of Electrical Communication,
             Tohoku University, Sendai, 980-8577, Japan}

\author{M. Lippmaa}
\affiliation{Institute for Solid State Physics, University of Tokyo,
             5-1-5, Kashiwanoha, Kashiwa, Chiba 277-8581, Japan}
             
\date{\today}

\begin{abstract}

We report on theoretical and experimental investigation of $A$-site
driven ferroelectricity in ferromagnetic La$_2$NiMnO$_6$ thin films
grown on SrTiO$_3$ substrates. Structural analysis and density
functional theory calculations show that epitaxial strain stretches the
rhombohedral La$_2$NiMnO$_6$ crystal lattice along the [111]$_\mathrm{cubic}$
direction, triggering a displacement of the $A$-site La ions in the
double perovskite lattice. The lattice distortion and the $A$-site
displacements stabilize a ferroelectric polar state in
ferromagnetic La$_2$NiMnO$_6$ crystals. 
The ferroelectric state only appears in the rhombohedral La$_2$NiMnO$_6$ phase,
where MnO$_6$ and NiO$_6$ octahedral tilting is inhibited by the 3-fold crystal symmetry.  
Electron localization mapping showed 
that covalent bonding with oxygen and 6$s$ orbital lone pair formation
are negligible in this material.

\end{abstract}


\maketitle

\section{INTRODUCTION}

The likelihood of a symmetry-breaking ferroelectric atomic displacement occurring
at the $B$-site of an $AB$O$_3$ perovskite can generally be predicted by evaluating the
Goldschmidt tolerance factor, $t=(r_\mathrm{O}+r_A)/\sqrt{2}(r_\mathrm{O}+r_B)$,
where $r_\mathrm{O}$, $r_A$, and $r_B$ are the ionic radii of the oxygen anion,
and the $A$- and $B$-site cations \cite{bilc_prl, benedek_jpcc}. 
When $t>1$, the $B$-site cation in the perovskite lattice has sufficient 
space to be displaced from the center of a $B$O$_6$ octahedron. 
Conversely, if $t<1$, an $A$-site displacement would be preferred 
over a $B$-site shift, as has been predicted theoretically for
K$_{0.5}$Li$_{0.5}$NbO$_3$ \cite{bilc_prl} and LaLuNiMnO$_6$ \cite{singh_prl}. 
Experimentally, Ba$_{1-x}$Ca$_x$TiO$_3$ ($0.02<x<0.34$) single 
crystals have been shown to exhibit off-center displacements
of the smaller Ca ions at the $A$-site \cite{fu_prl}.
However, smaller $A$-site ions generally lead to rotations and tilting 
of the $B$O$_6$ octahedra, which preserves the inversion symmetry by doubling the
unit cell \cite{bilc_prl, benedek_jpcc}.  
For this reason, $A$-site driven proper ferroelectricity is usually negligible in perovskites. 

Well-known exceptions are the Bi- and Pb-containing $A$-site ferroelectric perovskites 
Pb(Ti,Zr)O$_3$, BiFeO$_3$, and Bi$_2$NiMnO$_6$ 
\cite{cohen_n, wang_s, hill_jpcb, neaton_prb, azuma_jacs}. 
The Pb and Bi ions in those compounds form covalent bonds with oxygen and contain $6s$
orbital lone pairs, stabilizing a distorted structure and breaking the inversion symmetry. 
For other $A$-site ions, like Ba and La, 
it is exceedingly rare to form a covalent bond with oxygen 
and $6s$ orbital lone pairs in the perovskite lattice. 
Exceptions can be found in short-period PbTiO$_3$/SrTiO$_3$
$A$-site ordered superlattices \cite{bousquet_n} and
in $A$-site ordered double perovskites \cite{rondinelli_am,zhao_nc}. 
A combination of $A$-site ordering and $B$O$_6$ octahedral rotation in 
epitaxially-strained heterostructures has been shown to give rise to improper ferroelectricity. 
   
In this work, we have investigated the possibility of $A$-site driven ferroelectricity
in $B$-site ordered ferromagnetic La$_2$NiMnO$_6$ crystals. 
Based on Kanamori-Goodenough rules\cite{goodenough_pr,kanamori_jpcs}, 
La$_2$NiMnO$_6$ has been theoretically predicted and 
experimentally reported to show ferromagnetic order 
due to the presence of 180$^\circ$ Ni$^{2+}$-O-Mn$^{4+}$ superexchange bonding 
between an empty Mn$^{4+}$ $e_g$ orbital and a half-filled $d$ orbital on a 
neighboring Ni$^{2+}$ site. 
The average tolerance factor of La$_2$NiMnO$_6$ is 0.97, 
suggesting the possibility of $A$-site driven ferroelectricity. 
However, in bulk crystals, the $B$O$_6$ octahedra are tilted and rotated, 
forming non-polar rhombohedral (R$\overline{3}$) and monoclinic (P2$_1$/n) 
structures \cite{rogado_am, bull_jpcm, das_prl}.
  
La$_2$NiMnO$_6$ thin films can be grown epitaxially on 
LaAlO$_3$, (LaAlO$_3$)$_{0.3}$-(SrAl$_{0.5}$Ta$_{0.5}$O$_3$)$_{0.7}$ (LSAT), 
and SrTiO$_3$ substrates \cite{hashisaka_apl}. Epitaxial strain always shrinks or expands the 
crystal lattice of a thin film along a certain direction, 
possibly stabilizing a phase that is different from a strain-free 
bulk crystal. Even when no strain-related symmetry changes occur in
the film, it is common to see Curie temperature shifts in ferroelectrics 
and ferromagnets \cite{schlom_mrsb}. 
For example, epitaxial strain has been shown to induce ferroelectricity
in paraelectric SrTiO$_3$ \cite{pertsev_prb_2000}, SrMnO$_3$ \cite{lee_prl_2010}
and EuTiO$_3$ \cite{lee_n_2010}. In the case of La$_2$NiMnO$_6$ films 
on LaAlO$_3$(001) substrates, 
 polarized Raman measurements have revealed that the rhombohedral (R$\overline{3}$) 
 and monoclinic (P2$_1$/n) phases coexist at room temperature  \cite{iliev_apl}, 
 even though in bulk La$_2$NiMnO$_6$ crystals the rhombohedral 
 and monoclinic phases are stabilized at high and low temperatures, respectively
 \cite{rogado_am, bull_jpcm}. 
Here, we theoretically and experimentally investigate the epitaxial strain effect
on the dielectric properties of La$_2$NiMnO$_6$ thin films
and show that epitaxial strain induces ferroelectricity
in La$_2$NiMnO$_6$/SrTiO$_3$ heterostructures.  

\section{First-principles density functional calculations}

\begin{figure}
\includegraphics{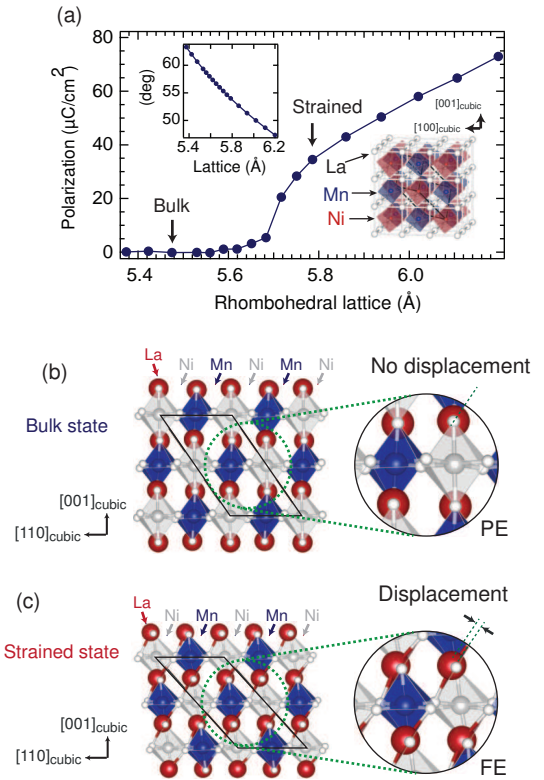}
\caption{(Color online) (a) Calculated strain dependence of spontaneous
 polarization along the [111]$_\mathrm{cubic}$ direction in a rhombohedral La$_2$NiMnO$_6$ lattice.
 The arrows in the graph mark the experimental bulk ($5.47$~\AA) and strained film
 lattice constants ($5.79$~\AA). The relationship between the rhombohedral lattice parameter
 and the rhombohedral angle $\alpha$ is shown in the left inset.
The right inset schematically illustrates the rhombohedral La$_2$NiMnO$_6$ crystal structure.
 The $B$-site Ni (red) and Mn (blue) atoms are alternately aligned along the
 [111]$_\mathrm{cubic}$ direction. 
 The black lines mark the rhombohedral unit cell. Schematic illustration of the
 paraelectric (b) and ferroelectric crystal structure under 5.9\% tension (c). 
 The black solid lines denote the rhombohedral La$_2$NiMnO$_6$ unit cell. 
 The paraelectric phase exhibits no displacement of the $A$-site La ions (red). 
 In contrast, a displacement of $A$-site La ions is clearly visible along the
 [111]$_\mathrm{cubic}$ direction
 in the ferroelectric phase of the strained rhombohedral La$_2$NiMnO$_6$ lattice. }
\label{calc}

\end{figure}

The effect of epitaxial strain on the crystal symmetry and polar instability
of La$_2$NiMnO$_6$ films was studied by density functional theory (DFT) simulations
using the Vienna Ab-initio Simulation Package (VASP) within the 
generalized gradient approximation to the exchange correlation potential \cite{kresse_prl, perdew_prl}. 
The electronic correlation effects were considered 
by using the generalized gradient approximation (GGA)+$U$ method 
with $U = 3$~eV for the Mn and Ni $d$ states \cite{zhu_apl}. 
It has been reported that La$_2$NiMnO$_6$ crystalizes 
in the rhombohedral structure (R$\overline{3}$) at high temperature and 
transforms to the monoclinic structure (P2$_1$/n) at low temperature. 
In fact, these two structures coexist over a wide temperature range \cite{rogado_am, bull_jpcm}. 
The strain effect on the polar instability was studied for both structures. 
For the rhombohedral structure, 
the unit cell volume was constrained to the experimentally observed value,
$V=353.33$~\AA$^3$, and converted to a hexagonal structure for the calculation. 
The $c/a$ ratio was tuned to simulate the strain effect. 
The cutoff energy for the plane waves was set at 400~eV, 
whereas the $k$-point sampling was done using a $4\times4\times2$ grid
in the hexagonal lattice. 
The Mn and Ni atoms were aligned alternately 
along the [111]$_\mathrm{cubic}$ direction with the spins set in a ferromagnetic configuration. 
The ferroelectric polarization was calculated by using the Berry phase method 
and comparing the ferroelectric structure and the paraelectric (centrosymmetric) reference structure. 

\begin{table}
\caption{Energy-minimized structural parameters of 
the paraelectric (top) and the ferroelectric (bottom) La$_2$NiMnO$_6$ structure under 5.9\% tension.}
\label{lattice}

Paraelectric phase: R$\overline{3}$ \\
\begin{tabular}{cccccc}
\hline \hline
$a$ (\AA)	& $\alpha$ ($^{\circ}$) &	& $x$	& $y$	& $z$	 	\\
\hline 
5.475	& 60.671 		& La	& 0.243	&0.243	&0.243	\\     
 		&			& Ni	& 0		&0		&0		\\     
 		&			& Mn	&0.500	&0.500	&0.500	\\     
 		&			& O	&0.787	&0.701	&0.273	\\
\hline \hline
 		&			& 	& 		&		&		\\     
\end{tabular}

Ferroelectric phase: R3, 5.9\% tension\\
\begin{tabular}{cccccc}
\hline \hline
$a$ (\AA)	& $\alpha$ ($^{\circ}$) &	& $x$	& $y$	& $z$	 \\
\hline 
5.253	& 53.997 		& La	& 0.272	&0.272	&0.272	\\     
		& 	 		& La	& 0.773	&0.773	&0.773	\\     
		&			& Ni	& 0.006	&0.006	&0.006	\\     
 		&			& Mn	& 0.505	&0.505	&0.505	\\     
 		&			& O	& 0.217	&0.303	&0.749	\\     
 		&			& O	& 0.247	&0.794	&0.704	\\     
\hline \hline
\end{tabular}
\end{table}

The monoclinic structure did not show spontaneous polarization under any strain field, 
corroborating the recent work by Zhao et al. \cite{zhao_nc}. 
In contrast, the rhombohedral La$_2$NiMnO$_6$ crystal structure, 
schematically illustrated in the inset of Fig.~\ref{calc}(a), 
presented a clear polar state under moderate strain. 
Fig.~\ref{calc}(a) shows the strain dependence of the spontaneous polarization along the
[111]$_\mathrm{cubic}$ direction in a rhombohedral La$_2$NiMnO$_6$ crystal. 
The inset plot shows the relationship between the rhombohedral 
lattice parameter and the rhombohedral angle. 
Increasing the tensile strain lowers the rhombohedral 
angle of the La$_2$NiMnO$_6$ crystal and 
ultimately leads to a polar state when the lattice parameter exceeds $5.7$~\AA, 
which corresponds to 4.2\% tension with respect to the experimental bulk value 
of $5.47$~\AA. A comparison of the partial density of states plots (not shown) of the 
paraelectric and a 5.9\% stretched ($\it{a}$=$5.79$~\AA) ferroelectric phases 
indicated that the empty La $6s$ state is very slightly shifted by the structural distortion, 
whereas the Ni and Mn $3d$ states remain essentially unaffected. 

\begin{figure}
\includegraphics{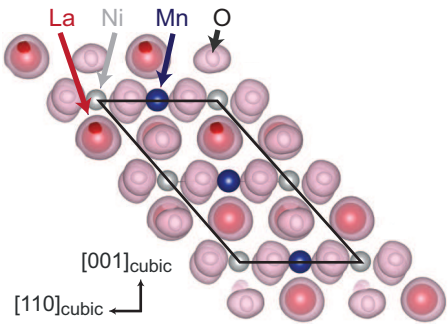}
\caption{Calculated electron localization functions for the ferroelectric R3 structure of
La$_2$NiMnO$_6$. The La ions are shown in dark red, the Ni ions in gray, the Mn ions in dark blue, 
the oxygen in white and the electron localization in pink.
The black lines mark the rhombohedral unit cell. }
\label{density}
\end{figure}

The crystal structures of the unstrained paraelectric
and the +5.9\% stretched ferroelectric rhombohedral La$_2$NiMnO$_6$ phases
are compared in Figs.~\ref{calc}(b) and \ref{calc}(c).
The structural model plots have been oriented along the [110]$_\mathrm{cubic}$ and
[001]$_\mathrm{cubic}$ directions. 
The $B$-site Ni and Mn ions are alternately aligned along the [111]$_\mathrm{cubic}$ direction. 
The relaxed structural parameters of the paraelectric and ferroelectric
La$_2$NiMnO$_6$ phases are listed in Table~\ref{lattice}.
In the paraelectric R$\overline{3}$ structure, the $A$-site La lattice shows no displacement. 
By applying a tensile strain field, the crystal symmetry drops from R$\overline{3}$ to R3 and 
the La lattice develops a sizable La displacement along the [111]$_\mathrm{cubic}$ direction. 
The mechanism responsible for the La lattice shift is fundamentally different from 
the Bi lone pair mechanism seen in ferroelectric BiFeO$_3$ \cite{wang_s, hill_jpcb, neaton_prb} and 
Bi$_2$NiMnO$_6$ \cite{azuma_jacs}.
The calculated electronic localization
function \cite{silvi_n} is mapped in Fig.~\ref{density} for the R3 symmetry of
ferroelectric La$_2$NiMnO$_6$. 
No unbalanced charge distributions were found around the La ions, 
unlike the ferroelectric Bi perovskites that have been reported to exhibit 
localized lobe-shaped charge distributions surrounding the Bi ions \cite{baettig_cm}.    
This indicates that the strain-induced ferroelectricity
in rhombohedral La$_2$NiMnO$_6$ cannot be attributed to 
either a 6$s$ orbital lone pair of La or covalent bonding between La and oxygen.
The DFT calculation suggests that ferroelectricity and ferromagnetism 
coexist in the rhombohedral La$_2$NiMnO$_6$ lattice when the crystal is
stretched along the [111]$_\mathrm{cubic}$ direction.

\section{EXPERIMENT}

To verify the DFT calculation results experimentally, La$_2$NiMnO$_6$ films 
were grown on SrTiO$_3$ ($c=3.905$~\AA) and LSAT ($c=3.87$~\AA) substrates by
pulsed laser deposition (PLD) \cite{kitamura_apl1, kitamura_apl2, sakurai_jap}. 
A polycrystalline stoichiometric La$_2$NiMnO$_6$ target, 
fabricated by conventional solid state reactions, was ablated with a KrF 
excimer laser ($\lambda=248$~nm) at a repetition rate of 5~Hz. 
During deposition, the SrTiO$_3$ substrates were kept at temperatures between 
600 and 700$^\circ$C and pure oxygen gas 
was continuously supplied into the growth chamber to maintain an ambient pressure of 500~mTorr. 
After growth, the films were cooled to 500$^\circ$C at a rate of 15$^\circ$C/min. 
The growth chamber was then filled to 760~Torr of 
pure oxygen gas and the films were cooled down to room temperature. 

Basic room-temperature structural analysis was done  
by x-ray diffraction (XRD) and reciprocal space mapping. 
As it is difficult to determine the crystal symmetry of 
La$_2$NiMnO$_6$ films by XRD, Raman spectroscopy was used to
determine the presence of different crystallographic phases and to look for 
structure changes below room temperature. 
A He-Ne laser (633~nm, 17~mW) was focused on the
La$_2$NiMnO$_6$ film surface through a $\times50$ objective lens with N.A.=1.0
for room-temperature measurements and 0.5 for low-temperature measurements. 
The scattering spectra were collected by a charge-coupled 
device detector (RAMASCOPE, Renishaw). 
As the film surface was parallel to the (001)$_\mathrm{cubic}$ plane, 
polarized Raman spectra could be taken in exact 
$XX$, $X'X'$, $XY$ and $X'Y'$ scattering configurations, 
where $X\parallel[110]_\mathrm{cubic}$, $X'\parallel[100]_\mathrm{cubic}$, 
$Y\parallel[\overline{1}10]_\mathrm{cubic}$, and $Y'\parallel[010]_\mathrm{cubic}$. 
The sample temperature was controlled with a He flow
cryostat (Microstat, Oxford Instruments). 

The degree of $B$-site order in La$_2$NiMnO$_6$ films can be estimated
from the saturation magnetization.
The magnetic properties of the films were measured 
with a superconducting quantum interference device (SQUID) magnetometer.

The presence of ferroelectric order was measured by several complementary
techniques to show that the ferroelectricity is indeed an intrinsic feature
of the films and not an artefact of temperature- or bias-dependent leak current
or interface capacitance changes.
For traditional ferroelectric $P-E$ hysteresis measurements, 
interdigitated Au electrodes were formed on the La$_2$NiMnO$_6$
film surfaces by e-beam evaporation and photolithographic lift-off \cite{takahashi_jap_2015}. 
The Au electrode thickness was 100~nm, the finger width and 
length were 10~$\mu$m and 200~$\mu$m,
respectively, and the electrode gap was 10~$\mu$m. 
Electrical contacts were made by ultrasonic wire bonding of 
aluminum wires to the Au pads. 
The electrical polarization was deduced by integrating the displacement current 
during a bipolar 15~V triangular bias voltage sweep at 1~Hz. 

Positive-Up-Negative-Down (PUND) measurements were performed for detecting 
the displacement current during ferroelectric domain reversal by an applied bias voltage
and eliminate the possibility that the $P-E$ loop measurements might be affected
by a bias-dependent interface capacitance change.
At first, all ferroelectric domains were poled by a trapezoidal 20~V pulse. 
Subsequently, two positive and two negative trapezoidal pulses 
were applied to the La$_2$NiMnO$_6$ films. 
The displacement current was converted to a voltage signal 
with a current-voltage converter and measured with a digitizer. 

Determining the temperature dependence of polarization from $P-E$
measurements can be complicated in thin films by changes in the magnitude
of leak currents that may completely mask small displacement currents.
A useful technique for measuring the polarization
of a thin film sample without having to apply a bias voltage during the
measurement is the Chynoweth method of dynamic pyroelectric detection
of dielectric polarization.
For the pyroelectric measurements \cite{takahashi_jap_2012,takahashi_jap_2015,
 takahashi_prb, takahashi_apl}, 
a 100-nm thick Pd top electrode was 
deposited on the La$_2$NiMnO$_6$/Nb:SrTiO$_3$ film surface by electron beam evaporation 
through a stencil mask with 1 mm diameter openings. 
Chopped light from a semiconductor diode laser (1.63~$\mu$m, 100~mW) 
was focused on a Pd top electrode pad, resulting in a modulation of the 
La$_2$NiMnO$_6$ capacitor temperature and the generation of a pyroelectric current. 
The sample current was converted to a voltage signal with a 
current-voltage converter and measured with a lock-in amplifier. 
The Pd electrode thickness of 100~nm was large enough to 
eliminate the occurrence of photocurrents under the 1.63~$\mu$m laser illumination. 
Details of the pyroelectric measurements can be found in
Refs.~\citenum{takahashi_jap_2012, takahashi_jap_2015, takahashi_prb, takahashi_apl}.    

Since polar and non-polar crystallographic phases are known to coexist in La$_2$NiMnO$_6$,
a scanning nonlinear dielectric microscope (SNDM) \cite{cho_rsi, cho_jmr} 
was used to image the nanoscale spatial distribution of 
ferroelectric polar domains at the La$_2$NiMnO$_6$ film surface 
at room temperature. 
A metal-coated cantilever attached to a 0.8~GHz LC oscillator
 SNDM probe was used to scan the film surface. 
A 1~V$_{pp}$, 23~kHz ac bias was applied between the 
Nb:SrTiO$_3$ substrate and the probe tip.       

\section{RESULTS AND DISCUSSION}

\subsection{Structural and magnetic properties of La$_2$NiMnO$_6$ films}

\begin{figure}
\includegraphics{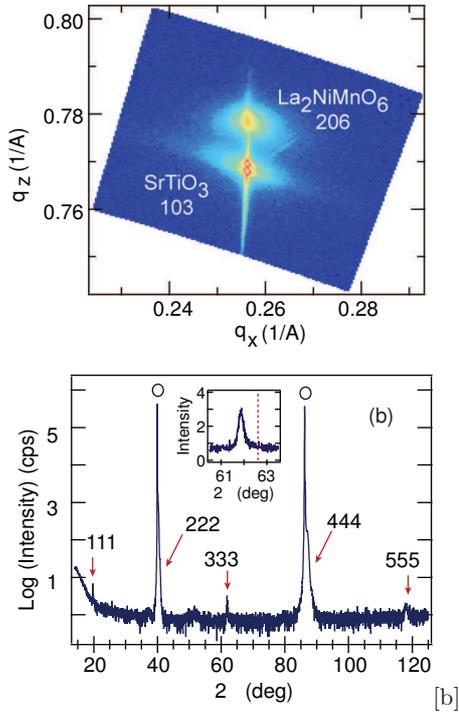}[b]
\caption{(Color online) (a) XRD reciprocal space map around the SrTiO$_3$ 
substrate (103)$_\mathrm{cubic}$ reflection of a La$_2$NiMnO$_6$/SrTiO$_3$ film, 
confirming coherent growth of the La$_2$NiMnO$_6$ film on the substrate. 
(b) XRD pattern of the La$_2$NiMnO$_6$ film along the SrTiO$_3$
substrate [111]$_\mathrm{cubic}$ direction,
indicating that the $B$-site Ni and Mn ions are ordered. 
Details of the (333)$_\mathrm{cubic}$ reflection are shown in the inset. 
The red line marks the expected bulk peak position, 
indicating that the La$_2$NiMnO$_6$ film crystal is stretched by +1.1\% 
along the [111]$_\mathrm{cubic}$ direction by epitaxial strain. }
\label{XRD}
\end{figure}

The degree of spontaneous ordering of $B$-site ions in double perovskite thin 
films can be tuned by varying the film growth temperature and the oxygen pressure, 
leading to various different magnetic phases, including ferromagnetic, ferrimagnetic, and antiferromagnetic films \cite{kitamura_apl1,ohtomo_jmr}. 
In this work, ordered and disordered La$_2$NiMnO$_6$ films were 
grown on SrTiO$_3$(001) substrates at 700$^\circ$C and 600$^\circ$C, respectively. 
In Fig.~\ref{XRD}(a), a reciprocal space map of a La$_2$NiMnO$_6$/SrTiO$_3$ sample grown 
at 700$^\circ$C shows a single (206) film reflection in addition to the strong substrate (103) peak. 
All samples used in this study presented similar coherent growth of 
La$_2$NiMnO$_6$ on both SrTiO$_3$(001) and LSAT(001) substrates. 
The in-plane and out-of-plane lattice parameters of the strained 
La$_2$NiMnO$_6$/SrTiO$_3$ film were $3.91$~\AA\ and $3.85$~\AA, respectively. 
In contrast, the lattice parameters of a La$_2$NiMnO$_6$/LSAT sample 
were $3.87$~\AA\ and $3.88$~\AA\ along the in-plane and out-of-plane directions. 
The SrTiO$_3$ substrate thus induced tensile strain, 
stretching the La$_2$NiMnO$_6$ lattice along the in-plane direction. 
The ordering of the $B$-site Ni and Mn ions was determined from 
$\theta/2\theta$ XRD measurements performed with a four-circle diffractometer
along the [111]$_\mathrm{cubic}$ direction, as shown in Fig.~\ref{XRD}(b). 
In addition to the SrTiO$_3$ $(lll)$ substrate peaks, 
all La$_2$NiMnO$_6$ $(lll)$ reflections can be clearly seen. 
Small peaks of odd-$l$ La$_2$NiMnO$_6$ $(lll)$ reflections 
correspond to the ordered $B$-site superlattice 
of Ni and Mn ions along the [111]$_\mathrm{cubic}$ direction. 
The inset shows details of the La$_2$NiMnO$_6$(333) reflection.
The red line marks the expected peak position of a non-strained bulk La$_2$NiMnO$_6$ crystal,
showing that the $B$-site ordered La$_2$NiMnO$_6$ film 
on the SrTiO$_3$ substrate was stretched by +1.1\% 
along the [111]$_\mathrm{cubic}$ direction by the epitaxial strain. 

\begin{figure*} 
\includegraphics{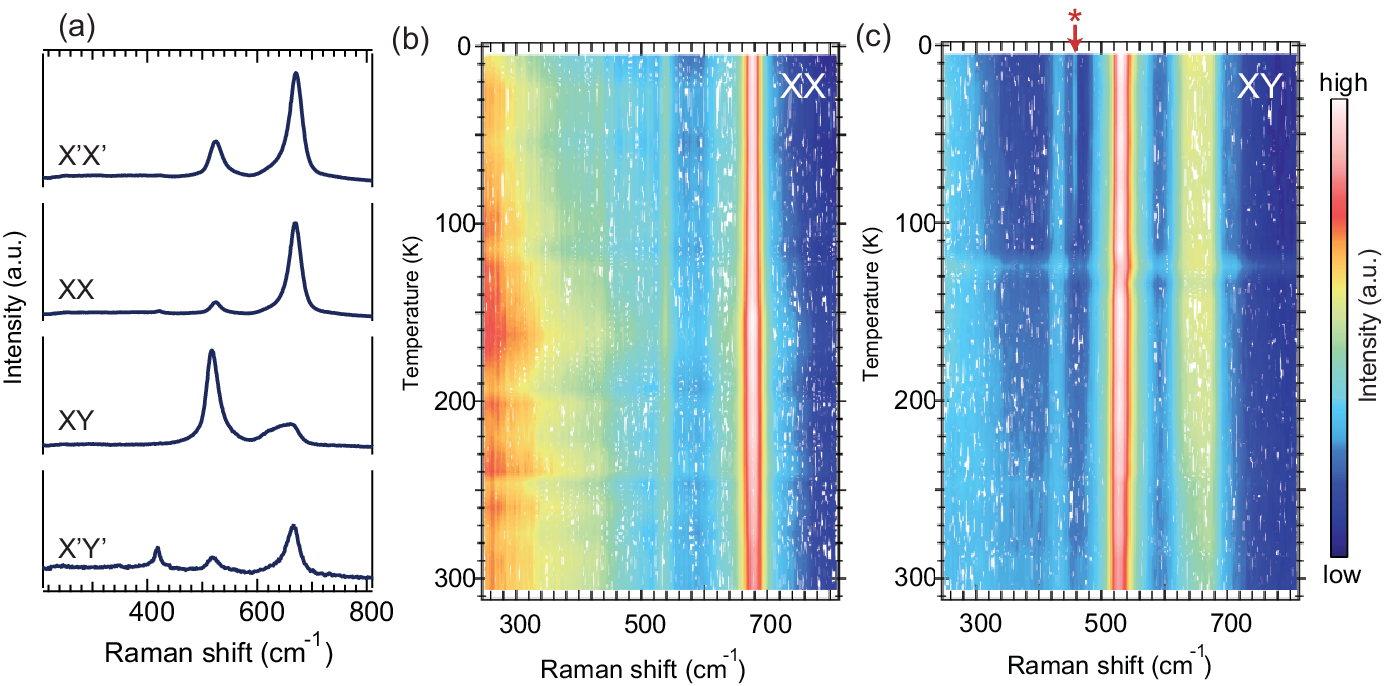}
\caption{(Color online) (a) Polarized Raman spectra of an ordered La$_2$NiMnO$_6$/SrTiO$_3$ film 
obtained at room temperature, showing that rhombohedral and monoclinic phases coexist.  
Temperature dependence of $XX$ (b) and $XY$ (c) Raman spectra 
of a $B$-site ordered La$_2$NiMnO$_6$/SrTiO$_3$ film between 5~K and 300~K. 
The red arrow in (c) marks the 457~cm$^{-1}$ peak that appears below 105~K.  }
\label{Raman}
\end{figure*}

Rhombohedral and monoclinic phases have been reported to coexist
in La$_2$NiMnO$_6$ at room temperature \cite{iliev_apl}. 
Even in a phase-pure rhombohedral La$_2$NiMnO$_6$ film grown on a
SrTiO$_3$(001) substrate, four reflection peaks should be visible
in a reciprocal space map around the SrTiO$_3$(103) reflection. 
The overlapping diffraction peaks of multiple domains of two possible
phases make detailed structural analysis difficult with a laboratory x-ray source.
Due to this, the reciprocal space map Fig.~\ref{XRD}(a) shows only a
single peak originating from the La$_2$NiMnO$_6$ film.
In order to accurately analyze the La$_2$NiMnO$_6$ film structure, 
polarized Raman measurements were performed
on the ordered La$_2$NiMnO$_6$/SrTiO$_3$ films at room temperature,
as shown in Fig.~\ref{Raman}(a). The main benefit of the Raman
measurement is the ability to distinguish between the rhombohedral
and monoclinic phases in the film.

Except for small differences in phonon line parameters, 
the spectra and their variation with scattering configuration closely resemble 
those reported for La$_2$NiMnO$_6$/LaAlO$_3$(001) films \cite{iliev_apl},
indicating that rhombohedral and monoclinic phases coexist in 
our La$_2$NiMnO$_6$/SrTiO$_3$ films as well. 
In particular, the presence of three peaks in a $X'Y'$
spectrum can be used to detect the presence of the rhombohedral phase, 
because the monoclinic phase should not exhibit any $X'Y'$ scattering peaks.  
 
In order to investigate the stability of the rhombohedral and monoclinic phases
at low temperature, the temperature dependence of $XX$ and $XY$ Raman spectra was
investigated from 5~K to 300~K, as shown in Figs.~\ref{Raman}(b) and \ref{Raman}(c).  
The spectra show that both rhombohedral and monoclinic phases coexist over
the whole temperature range.

In general, the Raman peak position ($\omega_\mathrm{anh}$) 
has a temperature dependence of
$\omega_\mathrm{anh}(T)=\omega_0-C(1+\frac{2}{e^{\hbar\omega_0/kT}-1})$, with
$\omega_0$ and $C$ being adjustable parameters. 
The Raman peak positions should thus shift to higher wavenumbers
as the temperature is decreased. 
In Fig.~\ref{Raman}(b) and (c), only the
peak at 660-700~cm$^{-1}$ in the $XX$ spectra shows a detectable temperature dependence,
shifting from 678.51~cm$^{-1}$ at 300~K to 676.73~$cm^{-1}$ at 5~K,
indicating anomalous phonon softening.
Similar softening caused by spin-phonon coupling
has been observed in other double perovskite systems,
such as La$_2$NiMnO$_6$/LaAlO$_3$ \cite{iliev_apl} 
and La$_2$CoMnO$_6$/SrTiO$_3$ \cite{iliev_prb}.
Spin-phonon coupling can stabilize a ferroelectric state in 
epitaxially strained ferromagnetic crystals, as happens in 
ferroelectric and ferromagnetic EuTiO$_3$ thin films 
grown on DyScO$_3$ substrates \cite{lee_n_2010, schlom_mrsb}.
A similar mechanism may be expected to trigger ferroelectricity in 
strained ferromagnetic La$_2$NiMnO$_6$ films.

Besides peak shifts, a small peak, marked by an arrow in Fig.~\ref{Raman}(c),
appeared at 457~cm$^{-1}$ in the $XY$ spectra below 105~K.
A similar low-temperature behavior was observed for a disordered La$_2$NiMnO$_6$/SrTiO$_3$ film. 
The SrTiO$_3$ substrate crystal undergoes an antiferrodistortive structural transition 
from a high-temperature cubic phase to a low-temperature tetragonal phase at 105~K, 
resulting in a 0.015\% in-plane lattice parameter change \cite{okazaki_mrb}. 
The epitaxial strain imposed on the film by the SrTiO$_3$ substrate thus changes slightly
at this temperature, resulting in a discontinuous temperature dependence of the Raman spectra 
in La$_2$NiMnO$_6$ samples. 
The reduction of the crystal symmetry of a La$_2$NiMnO$_6$ film
can therefore be expected to enhance the ferroelectric polarization at temperatures below 105~K.

\begin{figure}
\includegraphics{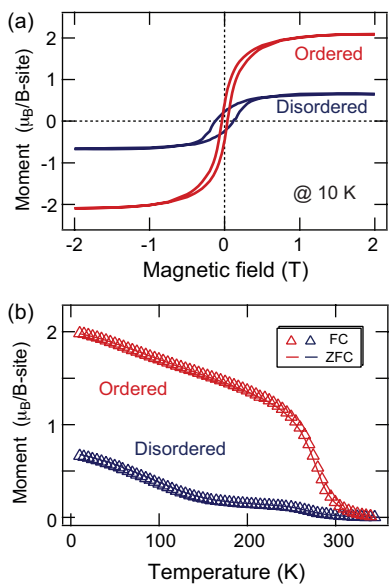}
\caption{(Color online) (a) Magnetization hysteresis loops at 10~K for ordered 
and disordered La$_2$NiMnO$_6$ films on SrTiO$_3$ substrates.
(b) Temperature dependence of the magnetic moment at 0.75~T, 
showing that the magnetic Curie temperature is close to 280~K, 
corresponding to the reported bulk value.}
\label{ferromagnetic} 
\end{figure}

The level of $B$-site order can be quantified by looking at the magnetization of the films, 
as shown in Fig.~\ref{ferromagnetic}(a). 
The magnetization hysteresis loops were measured at 10~K for ordered and 
disordered La$_2$NiMnO$_6$/SrTiO$_3$ samples. 
By comparing the observed saturation magnetizations of the films with the theoretical magnetization 
of 2.5~$\mu_{\mathrm B}$/$B$-site for a perfectly ordered La$_2$NiMnO$_6$ crystal 
\cite{goodenough_pr,kanamori_jpcs}, 
the fraction of B-site order in the ordered and disordered La$_2$NiMnO$_6$ films can be 
estimated to be 80\% and 26\%, respectively. 
Fig.~\ref{ferromagnetic}(b) presents the temperature dependence 
of the field-cooled and zero-field-cooled magnetizations at 0.75~T for the same samples. 
The magnetic Curie temperature is 280~K in both cases, matching reported values \cite{rogado_am}.
The magnetization data suggests that the ordered (80\%) and disordered (26\%) 
La$_2$NiMnO$_6$/SrTiO$_3$ samples were ferromagnetic and ferrimagnetic, respectively. 

\subsection{Ferroelectric properties of La$_2$NiMnO$_6$ films}

\begin{figure}
\includegraphics{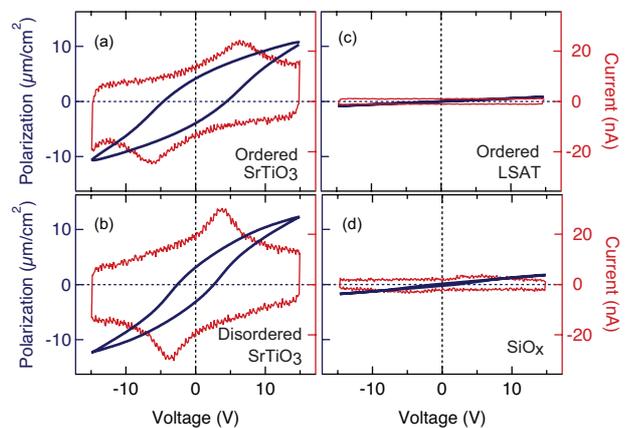}
\caption{(Color online) Ferroelectric polarization (left axis) and displacement current
(right axis) as a function of applied voltage at 10~K for ordered (a) and disordered (b)
La$_2$NiMnO$_6$/SrTiO$_3$ films, an ordered La$_2$NiMnO$_6$/LSAT film (c), and
a non-ferroelectric SiO$_x$/SrTiO$_3$ (d) reference sample.}
\label{hys} 
\end{figure}

The presence of spontaneous polarization was determined in ordered and disordered
La$_2$NiMnO$_6$/SrTiO$_3$ films, as well as an ordered La$_2$NiMnO$_6$/LSAT sample 
by measuring the ferroelectric polarization and the 
displacement current in the in-plane direction at 10~K as shown in Fig.~\ref{hys}. 
Both films grown on SrTiO$_3$ substrates were verified to exhibit ferroelectric domain reversal 
under an external electric field, while the La$_2$NiMnO$_6$/LSAT sample 
showed no hysteresis behavior at all, revealing the importance of strain 
in the formation of a ferroelectric state in the La$_2$NiMnO$_6$/SrTiO$_3$ films. 
There is little difference in ferroelectric polarization 
between the ordered and disordered La$_2$NiMnO$_6$/SrTiO$_3$ films. 
This is consistent with the calculation results in Fig.~\ref{calc}, 
which showed that the $B$-site Ni and Mn ions do not have 
a significant effect on the $A$-site driven ferroelectricity in La$_2$NiMnO$_6$ films. 
Moreover, this data can rule out improper ferroelectricity 
originating from bond- and site-centered charge ordering, 
as happens in Fe$_3$O$_4$ crystals \cite{brink_jpcm, takahashi_prb, takahashi_jap_2015} and 
the $E$-type antiferromagnetic order that can be found in 
Y$_2$CoMnO$_6$ and Lu$_2$CoMnO$_6$ crystals \cite{brink_jpcm, sharma_apl, vilar_prb}. 

We note that when measuring the ferroelectric polarization in thin films grown 
on SrTiO$_3$ substrates, it is important to verify that the substrate material 
does not contribute to the observed signal \cite{takahashi_jap_2015}. 
It has been reported that an electric field of 700~V/mm 
can induce a polarization of $\approx1$~$\mu$C/cm$^2$ in SrTiO$_3$ crystals  \cite{hemberger_prb}. 
To rule out the effect of the substrate on the La$_2$NiMnO$_6$ film measurements, 
a non-ferroelectric SiO$_x$ film with a thickness of 100~nm was grown 
on a SrTiO$_3$(001) substrate by thermal evaporation. 
After annealing the film at 400$^\circ$C in air and depositing identical 
interdigitated Au electrodes, ferroelectric hysteresis measurements were performed at 10~K. 
Fig.~\ref{hys}(d) shows the hysteresis loop and displacement current of a SiO$_x$/SrTiO$_3$ film 
for comparison with the ordered and disordered La$_2$NiMnO$_6$/SrTiO$_3$ films. 
The SiO$_x$/SrTiO$_3$ sample did not show a hysteresis loop, 
proving that the hysteresis loops obtained for the ordered and disordered 
La$_2$NiMnO$_6$/SrTiO$_3$ films originated from ferroelectricity in the films. 

\begin{figure}
\includegraphics{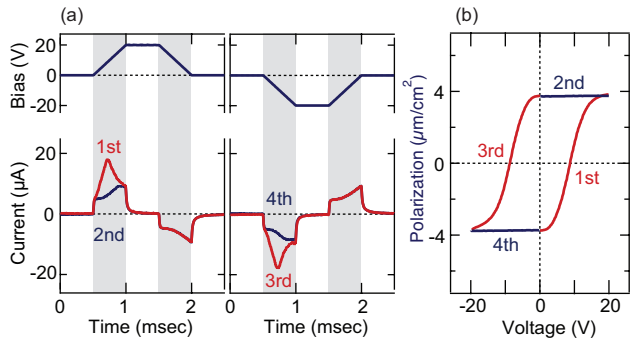}
\caption{(Color online) (a) PUND measurements at 10~K for an ordered
La$_2$NiMnO$_6$/SrTiO$_3$ sample. 
Only the first and third pulses showed a displacement current 
due to the reversal of ferroelectric domains. 
(b) Ferroelectric polarization hysteresis loop calculated 
from the PUND measurements in (a).} 
\label{PUND} 
\end{figure}

Successful observation of ferroelectric hysteresis loops
means that leakage currents in the La$_2$NiMnO$_6$ films were
low compared to the displacement current that arises when the ferroelectric
polarization is reversed by an applied bias. 
However, this type of direct hysteresis measurements 
is generally not a reliable method for measuring the spontaneous polarization
in ferroelectric oxide thin films because the  
measured current consists of several components due to switching 
of the ferroelectric domains, the linear dielectric response,
and leakage current of the material. 
Particularly for La$_2$NiMnO$_6$ films, the intermixed non-ferroelectric monoclinic phase
would produce a linear dielectric response when an electric field is applied,  
resulting in an increase of the background current. 
To suppress the linear dielectric behavior during hysteresis measurements, 
PUND measurements were performed for the ordered La$_2$NiMnO$_6$/SrTiO$_3$ sample,
as shown in Fig.~\ref{PUND}(a).

Only the first and third pulses showed domain reversal displacement currents, 
proving that the La$_2$NiMnO$_6$/SrTiO$_3$ film was indeed ferroelectric. 
The advantage of PUND measurements is to eliminate the linear dielectric current 
contribution by subtracting the second (fourth) cycle current from the first (third) cycle one.  
Fig.~\ref{PUND}(b) presents the ferroelectric hysteresis loop measured in this way, 
exhibiting a fully saturated loop with a spontaneous polarization of 3.8~$\mu$C/cm$^2$. 
The PUND measurement shows that the spontaneous polarization 
of the La$_2$NiMnO$_6$ layer is slightly lower than the remanent polarization 
seen in the standard hysteresis measurements in Fig.~\ref{hys}(a), which
include a linear dielectric response component.

\begin{figure}[!t]
\includegraphics{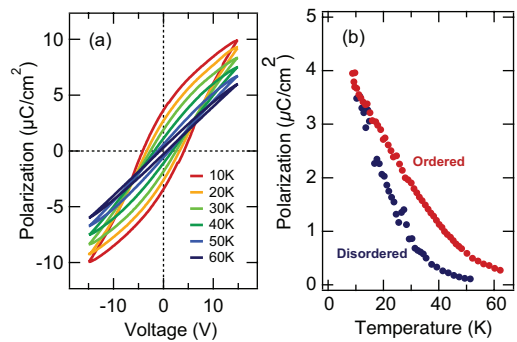}
\caption{(Color online) (a) Comparison of $P-E$ hysteresis loop shapes measured from 10 to 60~K. 
(b) Temperature dependence of the remanent polarization of ordered and
disordered La$_2$NiMnO$_6$ films.}
\label{hys_temp} 
\end{figure}

The temperature dependence of ferroelectric hysteresis loop shapes is presented
in Fig.~\ref{hys_temp}(a), together with the
temperature dependence of the remanent polarization in Fig.~\ref{hys_temp}(b). 
The remanent ferroelectric polarizations of ordered and disordered 
La$_2$NiMnO$_6$/SrTiO$_3$ films
were estimated to be 4.2~$\mu$C/cm$^2$ and 3.2~$\mu$C/cm$^2$, respectively, 
from the hysteresis loops measured at 10~K, and gradually decreased to 
zero at around 50 to 60~K as the temperature was increased. 

\begin{figure}
\includegraphics{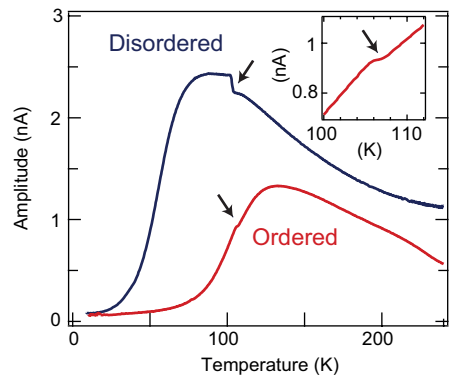}
\caption{(Color online) Temperature dependence of the pyroelectric response for
ordered (red) and disordered (blue) La$_2$NiMnO$_6$/SrTiO$_3$.
The inset shows the pyroelectric response of the ordered film sample close to 105~K. }
\label{pyro} 
\end{figure}

To precisely evaluate the polar state in La$_2$NiMnO$_6$/SrTiO$_3$ films, 
dynamic pyroelectric measurements \cite{takahashi_jap_2012, takahashi_jap_2015, takahashi_prb} 
were performed for the ordered and disordered 
La$_2$NiMnO$_6$/Nb:SrTiO$_3$(001) samples, as shown in Fig.~\ref{pyro}. 
In this case, the temperature derivative of the spontaneous polarization
was measured along the out-of-plane direction. 
Below 250~K, the pyroelectric response increased for both samples with decreasing temperature. 
The maximum pyroelectric signal amplitude was observed at 100 to 150~K, 
below which the pyroelectric response decreased. 
A clear kink can be seen in the pyroelectric current amplitude at 105~K, 
as marked by arrows in Fig.~\ref{pyro}. 
This is related to the abrupt change observed in the Raman measurements in Fig.~\ref{Raman}(c). 
The 105~K structural transition in the SrTiO$_3$ substrate crystal
causes a stepwise 0.015\% change in the epitaxial strain on the La$_2$NiMnO$_6$ film lattice,
resulting in a discontinuous temperature dependence of the pyroelectric responses in both samples. 
This result is a clear indication that ferroelectricity in 
La$_2$NiMnO$_6$/SrTiO$_3$ films is very sensitive to epitaxial strain. 

\begin{figure}
\includegraphics{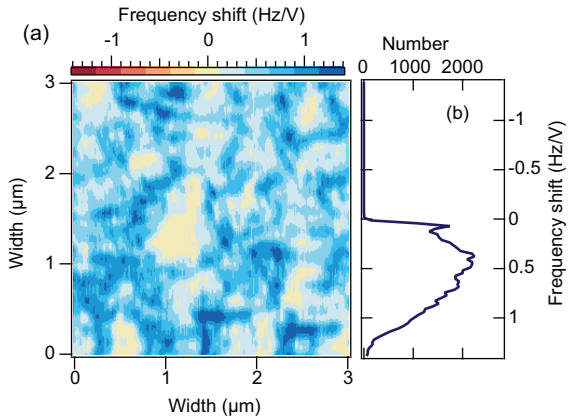}
\caption{(Color online) (a) SNDM image of an ordered La$_2$NiMnO$_6$/SrTiO$_3$ film surface
at room temperature. (b) The histogram of the SNDM signal reveals that most ferroelectric
domains are negatively polarized. Regions where the frequency shift is zero correspond
to the intermixed monoclinic phase.}
\label{SNDM} 
\end{figure}

The ferroelectric domain structure of the films was determined by acquiring 
SNDM \cite{cho_rsi, cho_jmr} images at room temperature. 
Fig.~\ref{SNDM}(a) shows a typical SNDM image 
for an ordered La$_2$NiMnO$_6$/Nb:SrTiO$_3$ film. 
The observation of a uniform SNDM frequency shift signal indicates 
that the La$_2$NiMnO$_6$/SrTiO$_3$ 
film was spontaneously polarized at room temperature. 
The histogram in Fig.~\ref{SNDM}(b) shows 
that most areas were polarized with a negative charge at the top surface, 
matching the polarization direction determined by the pyroelectric analysis. 
Some regions exhibited a zero frequency shift, indicating that there was no 
vertical spontaneous polarization in these parts of the La$_2$NiMnO$_6$/Nb:SrTiO$_3$ film,
probably due to the intermixing of the non-polar monoclinic phase that was detected
in the Raman measurements. 

According to the DFT calculations, a ferroelectric distortion appears only in 
the rhombohedral La$_2$NiMnO$_6$ phase 
when the crystal is stretched along the [111]$_\mathrm{cubic}$ direction. 
The monoclinic structure lacks the 3-fold rotation, 
allowing anti-phase tilting of the MnO$_6$ and NiO$_6$ octahedra, 
which suppresses the La polar distortion and results in a paraelectric state. 
In contrast, the rhombohedral structure preserves the 
3-fold rotation symmetry that does not allow MnO$_6$ and NiO$_6$ 
octahedral tilting but allows polar ionic shifts of La. 
Therefore, ferroelectricity is induced only in a strained 
La$_2$NiMnO$_6$/SrTiO$_3$ film and cannot appear in bulk La$_2$NiMnO$_6$. 

\section{CONCLUSION}

We have theoretically and experimentally demonstrated the presence of 
strain-induced ferroelectricity in ferromagnetic La$_2$NiMnO$_6$ films. 
Theoretical calculations and tolerance factor analysis indicated the presence  
of $A$-site driven ferroelectricity along the [111]$_\mathrm{cubic}$
direction in epitaxially strained La$_2$NiMnO$_6$ crystals. 
A polar state was shown to exist in epitaxial La$_2$NiMnO$_6$ films even at room temperature. 
The observation of $A$-site ferroelectricity in epitaxially-strained La$_2$NiMnO$_6$ films 
highlights the potential of strain engineering for tailoring the properties of 
complex oxides and extending their functionalities for new device applications. 

\begin{acknowledgments}

This study was partly supported by JSPS KAKENHI 
(Grant Nos. 26105002, 25706022, 24740235, and 23226008). 
This work was also partly supported by 
the Kurata Memorial Hitachi Science and Technology Foundation
and the Iketani Science and Technology Foundation. 
MK acknowledges the financial support from 
the Japan Society for the Promotion of Science for young scientists. 
The authors would like to thank N. Chinone 
for technical support in using the SNDM.  

\end{acknowledgments}

%

\end{document}